\documentclass[10pt]{iopart}
\usepackage{xcolor}
\usepackage{graphicx} 
\usepackage{siunitx}
\usepackage{isotope}
\usepackage[colorlinks=true]{hyperref}
\usepackage{upgreek}

\begin{document}

\title{Detector characterization for a new $^{12}$C+$^{12}$C reaction study at LUNA}

\author{R.\,M.~Gesu\`e$^{1,2}$, S.~Turkat$^{3,4,\ast}$, J.~Skowronski$^{3,4}$, M.~Aliotta$^5$, L.~Barbieri$^5$, F.~Barile$^6$, D.~Bemmerer$^7$, A.~Best$^{8,9}$, A.~Boeltzig$^7$, C.~Broggini$^4$, C.\,G.~Bruno$^5$, A.~Caciolli$^{3,4}$, M.~Campostrini$^{10}$, F.~Casaburo$^{11,12}$, F.~Cavanna$^{13}$, T.~Chillery$^{2}$, G.\,F.~Ciani$^{6,14}$, P.~Colombetti$^{13,15}$, A.~Compagnucci$^{1,2}$, P.~Corvisiero$^{11,12}$, L.~Csedreki$^{16}$, T.~Davinson$^5$, D.~Dell'Aquila$^{8,9}$, R.~Depalo$^{17,18}$, A.~Di~Leva$^{8,9}$, Z.~Elekes$^{16,19}$, F.~Ferraro$^{2,\dag}$, A.~Formicola$^{20}$, Zs.~Fülöp$^{16}$, G.~Gervino$^{13,15}$, A.~Guglielmetti$^{17,18}$, C.~Gustavino$^{20}$, Gy.~Gyürky$^{16}$, G.~Imbriani$^{8,9}$, M.~Junker$^2$, M.~Lugaro$^{21,22}$, P.~Marigo$^{3,4,\S}$, J.~Marsh$^5$, E.~Masha$^7$, R.~Menegazzo$^4$, D. Mercogliano$^{8,9}$, V.~Paticchio$^6$, D.~Piatti$^{3,4}$, P.~Prati$^{11,12}$, D.~Rapagnani$^{8,9}$, V.~Rigato$^{10}$, D.~Robb$^5$, L.~Russell$^2$, R.\,S.~Sidhu$^{5,23}$, B.~Spadavecchia$^2$, O.~Straniero$^{20,24}$, T.~Szücs$^{16}$, S.~Zavatarelli$^{11,12}$ (LUNA collaboration)}



\address{$^1$ Gran Sasso Science Institute, 67100 L'Aquila, Italy}
\address{$^2$ INFN, Laboratori Nazionali del Gran Sasso, 67100 Assergi, Italy}
\address{$^3$ Università degli Studi di Padova, 35131 Padova, Italy}
\address{$^4$ INFN, Sezione di Padova, 35131 Padova, Italy}
\address{$^5$ SUPA, School of Physics and Astronomy, University of Edinburgh, EH9 3FD Edinburgh, United Kingdom}
\address{$^6$ INFN, Sezione di Bari, 70125 Bari, Italy}
\address{$^7$ Helmholtz-Zentrum Dresden-Rossendorf, 01328 Dresden, Germany}
\address{$^8$ Universit\`a degli Studi di Napoli ``Federico II'', 80125 Naples, Italy}
\address{$^9$ INFN, Sezione di Napoli, 80125 Naples, Italy}
\address{$^{10}$ Laboratori Nazionali di Legnaro, 35020 Legnaro, Italy}
\address{$^{11}$ Università degli Studi di Genova, 16146 Genova, Italy}
\address{$^{12}$ INFN, Sezione di Genova, 16146 Genova, Italy}
\address{$^{13}$ INFN, Sezione di Torino, 10125 Torino, Italy}
\address{$^{14}$ Università degli Studi di Bari ``A. Moro'', 70125 Bari, Italy}
\address{$^{15}$ Universit\`a degli Studi di Torino, 10125 Torino, Italy}
\address{$^{16}$ HUN-REN Institute for Nuclear Research (HUN-REN ATOMKI), PO Box 51, H-4001 Debrecen, Hungary}
\address{$^{17}$ Università degli Studi di Milano, 20133 Milano, Italy}
\address{$^{18}$ INFN, Sezione di Milano, 20133 Milano, Italy}
\address{$^{19}$ Institute of Physics, Faculty of Science and Technology, University of Debrecen, H-4032 Debrecen, Hungary} 
\address{$^{20}$ INFN, Sezione di Roma, 00185 Roma, Italy}
\address{$^{21}$ Konkoly Observatory, Research Centre for Astronomy and Earth Sciences (CSFK), HUN-REN, and MTA Centre for Excellence, 1121 Budapest, Hungary}
\address{$^{22}$ ELTE E\"otv\"os Lor\'and University, Institute of Physics, 1117 Budapest, Hungary}
\address{$^{23}$ School of Mathematics and Physics, University of Surrey, Guildford, GU2 7XH, United Kingdom}
\address{$^{24}$ INAF-Osservatorio Astronomico d'Abruzzo, 64100, Teramo, Italy}

\begin{footnotesize}
~\\
$\ast$ steffen.turkat@tu-dresden.de, 
$\dag$ federico.ferraro@lngs.infn.it\\
$\S$ Deceased
\end{footnotesize}

\begin{abstract}

The $^{12}$C+$^{12}$C fusion reaction plays a crucial role in stellar evolution, including the occurrence of supernova explosions, and in the synthesis of the chemical elements. However, our understanding of its cross section remains severely deficient, particularly below $E_\textrm{cm}=2.5$\,MeV, the energy range of interest for astrophysics. To address these unresolved issues, the LUNA collaboration will conduct a dedicated study of the $^{12}$C+$^{12}$C reaction at the Bellotti Ion Beam Facility (Bellotti IBF) located deep underground within the Gran Sasso National Laboratory (LNGS) in Italy. Based on the combination of passive and active shields, this campaign aims to achieve unprecedented sensitivity in measuring the cross sections of the two key reaction channels, $^{12}$C($^{12}$C,$\alpha$)$^{20}$Ne and $^{12}$C($^{12}$C,$p$)$^{23}$Na in the low-energy regime via $\gamma$-ray detection. Here, we report on a sensitivity study for the upcoming campaign with a focus on the characterization of two detectors, namely a HPGe detector and a NaI(Tl) array. Furthermore, their intrinsic contamination is thoroughly investigated since this could potentially influence the overall sensitivity. Assuming typical beam intensities of the Bellotti IBF, we will be able to investigate reaction rates significantly below 100 counts per day. In case of the $^{12}$C+$^{12}$C reaction we therefore expect to acquire experimental data well below the current limit of $E_\textrm{cm}=2.1\,$MeV. The results are supported by simulations to highlight the advantageous low-background environment, essential for high-precision nuclear astrophysics studies. 

\end{abstract}

\ioptwocol

\section{The $^{12}$C+$^{12}$C campaign at the Bellotti IBF}

Carbon burning is a key stage of stellar evolution, whose understanding is of paramount importance for the comprehension of supernova (SN) occurrence and outcomes. SNs provide a major contribution to the chemical and physical evolution of galaxies, they are distance ladders used to probe the past history of the Universe, and they generate the most compact objects in nature, such as neutron stars and black holes. However, the large uncertainty affecting the low-energy cross section of the $^{12}$C+$^{12}$C reaction hampers the knowledge of the final fate of stars \cite{Chieffi2021,Mori2018,DeGeronimo2024,Dumont2024,Seong2024}. 
The $^{12}$C+$^{12}$C fusion in stars proceeds primarily through the $^{12}$C($^{12}$C,$\alpha$)$^{20}$Ne and the $^{12}$C($^{12}$C,$p$)$^{23}$Na reactions. Successful modeling of supernovae requires the reaction rate to be known below 2\,MeV, but due to the vanishing cross sections caused by the Coulomb barrier, experiments are extremely challenging.

Measurements can be performed by detecting the $\gamma$-rays produced in the decay of the excited states of $^{23}$Na and $^{20}$Ne \cite{High1977, Kettner1977, Kettner1980, Dasmahapatra1982, Aguilera2006, Barron2006, Spillane2007}, by identifying the charged particles ($p$ and $\alpha$) emitted in the two reactions \cite{Patterson1969, Mazarakis1973, Becker1981, Zickefoose2018, Morales2024}, and by correlating $\gamma$-rays with the corresponding emitted particles in coincidence measurements \cite{Jiang2018, Tan2020, Fruet2020}. Each technique has advantages and disadvantages: The detection of $\gamma$-rays in a highly reduced background environment currently offers the best way to reach energies below 2\,MeV, but is not able to investigate the total cross section due to the non-detectable ground state transition. Particle detection provides a full knowledge of the cross section, but suffers the difficulty to identify the emitted particles. Coincidence measurements between $\gamma$-rays and particles help in the identification at the cost of a reduced detection efficiency and the inability in quantifying the $\alpha_0$ and p$_0$ transitions.

The LUNA collaboration (Laboratory for Underground Nuclear Astrophysics) aims at the measurement of the $^{12}$C+$^{12}$C fusion cross section through $\gamma$-ray detection, exploiting the intense $^{12}$C beam provided by the 3.5\,MV accelerator of the Bellotti IBF located underground at the Gran Sasso National Laboratory \cite{Sen2019, Junker2023}.
Thanks to 1400 meters of rock overlying the experimental halls, LNGS is the ideal location to perform a $\gamma$-ray detection experiment. We will combine thick passive lead shielding together with the high resolution of a HPGe detector and the high efficiency of scintillating NaI(Tl) detectors. The latter can also work as Compton suppression, providing an extra background reduction for the HPGe detector. The proposed configuration will reduce the $\gamma$-ray background by more than four orders of magnitude with respect to previous attempts at the energies of interest \cite{Spillane2007}. In fact, thanks to the underground location, a thicker lead shielding results in further reduction of the background, as background from cosmic ray induced muons can be considered negligible with respect to the signal expected by the reaction.

The final goal is to measure with unprecedented precision and accuracy the $^{12}$C+$^{12}$C cross section at stellar energies and to investigate the impact of the experimental results on our understanding of the evolution and nucleosynthesis of SN progenitors.
In the present paper we discuss the characterization of the HPGe detector and present a background and sensitivity study for the upcoming $^{12}$C+$^{12}$C campaign (cf. \autoref{sec:GePD2}). While intrinsic contaminations are usually negligible with respect to the expected count rate, as well as the natural and the in-beam background, these contributions in fact require a thorough investigation for the upcoming campaign. Furthermore, we provide a comprehensive investigation of a NaI(Tl) prototype, which is supported by dedicated simulations to investigate intrinsic contaminations. The results are subsequently compared to other low-background NaI(Tl) detection arrays (cf. \autoref{sec:NaI}). In upcoming publications we will also report in more detail on the final setup configuration, as well as the impact of unlikely in-beam contributions, as expected from \cite{Spillane2007}.

\section{The GePD2 detector}
\label{sec:GePD2}
The HPGe detector that will be used for the upcoming $^{12}$C+$^{12}$C campaign at the Bellotti IBF, hereafter GePD2, is a coaxial p-type detector from AMETEK Ortec with a relative efficiency of 157\,\%. It is manufactured in a low-background configuration and is equipped with a 1.5\,mm-thick copper endcap (108\,mm diameter). The crystal itself has a nominal length of 94.8\,mm and diameter of 92.0\,mm.

\begin{figure}[tbh!]
\includegraphics[width=\columnwidth,trim=15 25 0 0mm]{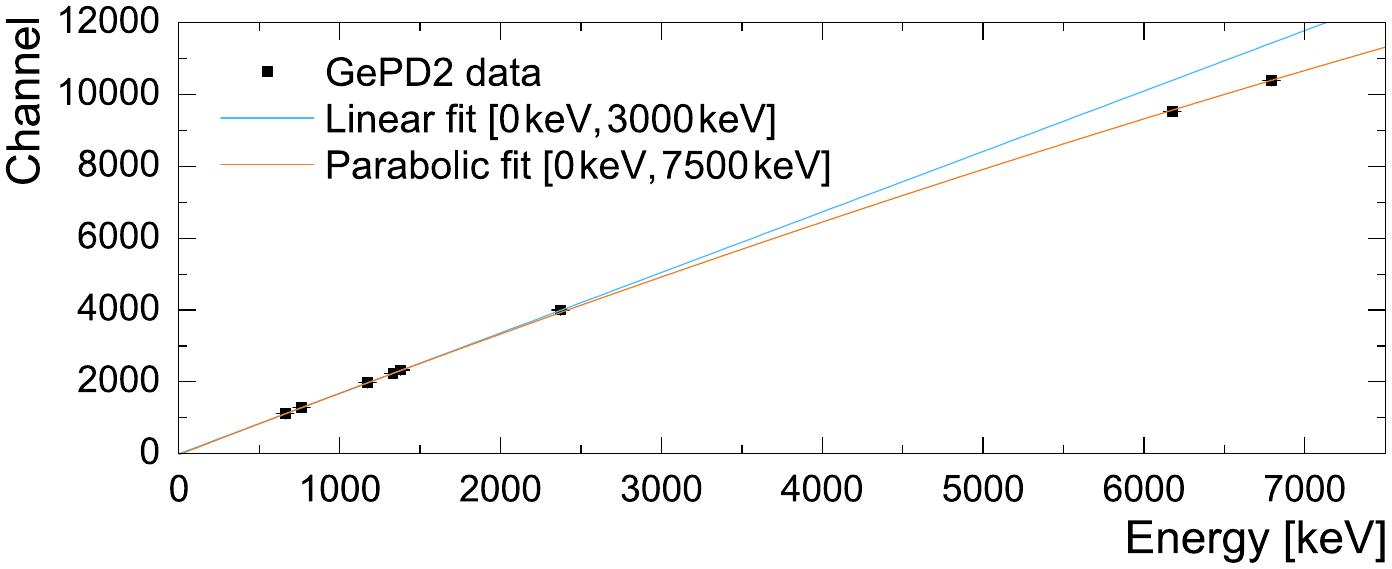}
\caption{\label{fig:GePD2_Ecal} Experimental energy calibration of GePD2 based on $^{60}$Co, $^{137}$Cs radioactive sources and $^{14}$N(p,$\gamma$)$^{15}$O reaction data. The entire data set is fitted with a second order polynomial (orange). In addition, a linear fit (blue) is used to solely represent the low-energy data and extrapolated to higher energies for comparability.}
\end{figure}

In case of comparatively low energies, i.e below 3\,MeV, the energy calibration of GePD2 can be well described by a linear function. This was investigated and confirmed using two available $^{60}$Co and $^{137}$Cs calibration sources. Based on additional data obtained via the $^{14}$N(p,$\gamma$)$^{15}$O reaction ($Q=7297$\,keV) at the Bellotti IBF, a parabolic fit is needed, as shown in \autoref{fig:GePD2_Ecal}. These non-linearities in energy calibrations are mainly associated to the electronic amplification in the signal processing chain \cite{Debertin1988, Knoll2010}. In contrast, any potential detector-intrinsic ballistic effects mainly affect the energy dependent resolution. This is due to high ionization densities leading to incomplete charge collection based on an increased probability of electron-hole recombination. The comparatively large crystal size of GePD2 further intensifies these ballistic deficits due to its impact on the overall charge collection time and the uniformity of the electric field within the crystal.

\subsection{Long-term background}

\begin{figure*}[tbh!]
\includegraphics[width=\textwidth,trim=8mm 5mm 15mm 0mm]{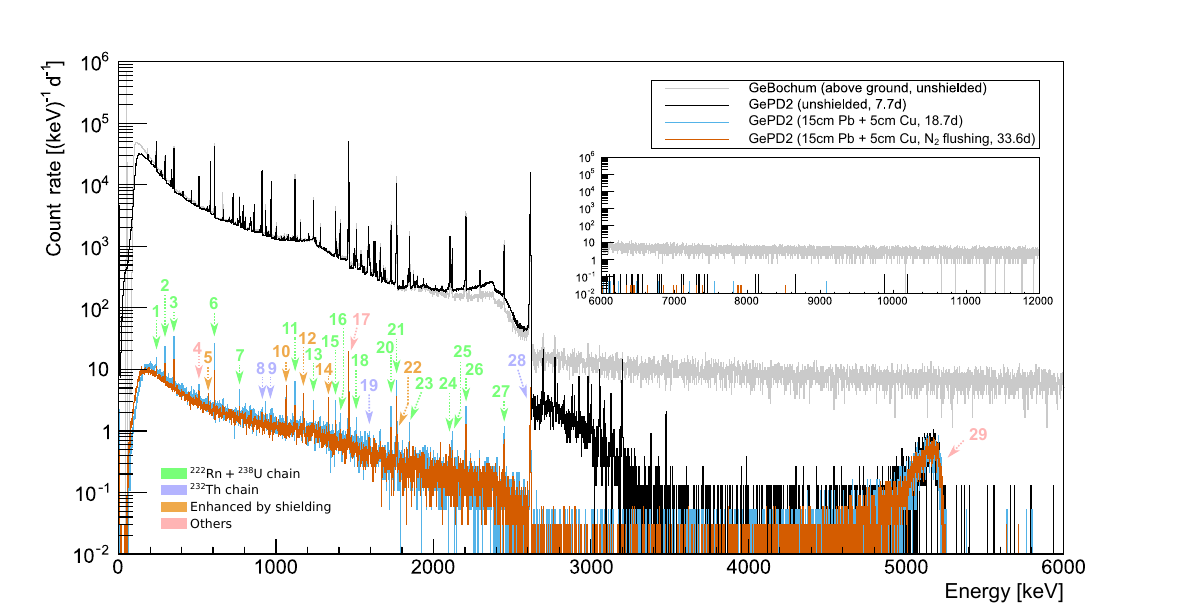}
\caption{\label{fig:GePD2_Spectrum} Comparison of the count rates for different shieldings of GePD2, i.e. without lead shielding (black), with a lead and copper shielding (blue) and an additional nitrogen flushing (orange). As a comparison, the count rate of the unshielded GeBochum above ground is shown in grey. The most prominent peak structures are labeled and color-coded according to their origin and listed in \autoref{tab:GePD2ReamainingPeaks}. The high energetic region from 6\,MeV to 12\,MeV is shown as an inset for the same spectra.}
\end{figure*}


In order to characterize the GePD2 detector and investigate its sensitivity for the upcoming $^{12}$C+$^{12}$C campaign at the Bellotti IBF, the detector was tested using different setups. First, the detector was operated without any additional shielding at the underground facility (cf. black curve in \autoref{fig:GePD2_Spectrum}). In addition, the detector was placed within a graded shield of 15\,cm of outer lead, as well as 5\,cm of inner copper to shield against the radioactivity stemming from the lead itself. Within this passive shielding GePD2 was operated both without (cf. blue curve in \autoref{fig:GePD2_Spectrum}) and with nitrogen flushing (cf. orange curve in \autoref{fig:GePD2_Spectrum}). This nitrogen flushing was implemented by guiding the radon-free nitrogen from the boil-off of the HPGe dewar into the lead castle, i.e. without the usage of an anti-radon box.

It was avoided to keep GePD2 in overground environments for longer measurements due to inevitable activation of long-living radionuclides by cosmic rays within the crystal, the cryostat, the cold finger and the detector holder, i.e. $^{22}$Na, $^{26}$Al, $^{54}$Mg, $^{57}$Co, $^{58}$Co, $^{60}$Co, $^{65}$Zn and $^{68}$Ge \cite{Heusser1993, Breier2020, Turkat2023}. Hence, the GePD2 spectra acquired underground are compared to the spectrum of another HPGe operated above ground called GeBochum instead (cf. grey curve in \autoref{fig:GePD2_Spectrum}), which is comparable in size and relative efficiency \cite{Boeltzig2018}.

In case of the unshielded GePD2 detector operated underground, the resulting pulse height spectra show typical background contributions mainly stemming from the natural decay chains of $^{238}$U and $^{232}$Th, as well as of $^{40}$K. Comparing this spectrum to typical spectra obtained in overground environments (grey curve) it is apparent that the $\gamma$-induced components are very similar, while the muon-induced continuum is strongly suppressed, as expected \cite{Mei2006}. This suppression also reveals low-intensity high energy $\gamma$-induced contributions between 2650\,keV and 3750\,keV stemming from $^{214}$Bi and $^{208}$Tl (black curve), which are clearly visible even down to emission probabilities of $\eta=0.0024(3)$\% in case of the $E=2826.96$\,keV line of $^{214}$Bi.

The shielded setups (blue and orange) show an attenuation in the count rate of more than three orders of magnitude across the entire region up to 2.6\,MeV compared to the unshielded case. All well-pronounced peaks are labeled (cf. \autoref{fig:GePD2_Spectrum}) and explained (cf. \autoref{tab:GePD2ReamainingPeaks}). Less pronounced peak structures, i.e. due to $\gamma$-rays from $^{214}$Bi with low emission probabilities, are not labeled due to their insignificance. In general, all visible structures are identified with the only exception of a peak at $E\approx 2042\,$keV. It's origin remains unresolved with the only reasonable explanation being a $\gamma$-ray from $^{194}$Au, fed by it's long-living mother nucleus $^{194}$Hg, as also discussed for the GATOR detector \cite{Angle2008}. There, it is reported to stem from a contamination in the surrounding lead bricks due to earlier activations.

While the spectrum of the setup without nitrogen flushing (blue) still shows pronounced contributions from $^{214}$Bi and other daughters from $^{222}$Rn, i.e. $^{214}$Pb, these can be further attenuated effectively with the installation of the nitrogen flushing. Comparing the ratio of the non-shielded and shielded spectra also allows for an investigation of the spatial origin of background contributions: In case of dominating background sources stemming from outside the lead shield, this ratio is expected to be a comparatively smooth curve without major peaks. There are however five peaks emerging from this analysis (labeled with 'Enhanced by shielding' in \autoref{fig:GePD2_Spectrum}). These peaks can be associated to $^{207}$Bi and $^{60}$Co. They are most likely less attenuated than the rest of the spectrum due to their presence within the lead bricks itself, rather than their presence within the GePD2 detector. Otherwise, this ratio analysis revealed no prominent peak structures, thus pointing towards a low contribution of intrinsic contaminations to the counting rate of the shielded spectrum, which therefore remains still predominantly influenced by external sources. 

It is also worthwhile mentioning that, among the identified peaks in the shielded spectra, none of them corresponds to the activation of long-living radionuclides (cf. \autoref{tab:GePD2ReamainingPeaks}), further emphasizing the importance of continuous underground operation. The only exception are again $^{60}$Co and $^{207}$Bi, which are most likely located within the utilized shielding itself. Furthermore, all acquired spectra of GePD2 are free of any additional structure above 4\,MeV besides one broad structure at 5.1\,MeV. This peak will be further discussed in \autoref{sec:GePD2IntrinsicAlpha}.

\begin{table}[tbh!]
\setlength{\tabcolsep}{2pt}
\caption{\label{tab:GePD2ReamainingPeaks}Visible background peaks in the shielded spectra of GePD2, as well as their origin. The numbering is according to the labels in \autoref{fig:GePD2_Spectrum} and 'SE' indicates single escape peaks.\vspace{3pt} 
}
\resizebox{\columnwidth}{!}{
\begin{tabular}{l|c|l||l|c|l||l|c|l}
$\#$ & \textrm{$E$\,[keV]} & \textrm{Origin} & $\#$ & \textrm{$E$\,[keV]} & \textrm{Origin} & $\#$ &\textrm{$E$\,[keV]} & \textrm{Origin}\\
\hline
1 & 242.0  & $^{214}$Pb & 11 & 1120.3 & $^{214}$Bi & 21 & 1764.5 & $^{214}$Bi    \\
2 & 295.2  & $^{214}$Pb & 12 & 1173.2 & $^{60}$Co  & 22 & 1770.2 & $^{207}$Bi    \\
3 & 351.9  & $^{214}$Pb & 13 & 1238.1 & $^{214}$Bi & 23 & 1847.4 & $^{214}$Bi    \\
4 & 511.0  & Annih & 14 & 1332.5 & $^{60}$Co  & 24 & 2103 (SE) & $^{208}$Tl \\
5 & 569.7  & $^{207}$Bi & 15 & 1377.7 & $^{214}$Bi & 25 & 2118.5 & $^{214}$Bi    \\
6 & 609.3  & $^{214}$Bi & 16 & 1408.0 & $^{214}$Bi & 26 & 2204.1 & $^{214}$Bi    \\
7 & 768.3  & $^{214}$Bi & 17 & 1460.8 & $^{40}$K   & 27 & 2447.7 & $^{214}$Bi    \\
8 & 911.2  & $^{228}$Ac & 18 & 1509.2 & $^{214}$Bi & 28 & 2614.5 & $^{208}$Tl    \\
9 & 969.0  & $^{228}$Ac & 19 & 1588.2 & $^{228}$Ac & 29 & 5304.3 & $^{210}$Po \\
10 & 1063.7 & $^{207}$Bi & 20 & 1729.6 & $^{214}$Bi & \\
\end{tabular}
}
\end{table}

The integrated count rates were determined within an energy interval of [200\,keV,2700\,keV] and result in $R=6\,034\,800\,(900)\,$d$^{-1}$ (without shielding), $R=4176(15)\,$d$^{-1}$ (with shielding and without nitrogen flushing), and $R=3416(10)\,$d$^{-1}$ (with shielding and nitrogen flushing), respectively. Assuming a crystal mass of approximately 3\,kg, the background count rate of GePD2 is $R\approx$1000\,kg$^{-1}$d$^{-1}$ with this preliminary setup.

The final setup for the upcoming $^{12}$C+$^{12}$C campaign will further increase the amount of surrounding lead by additional 5\,cm to a total of 20\,cm thickness. Based on the above discussions on the comparison between the shielded and non-shielded spectra, it is reasonable to assume that further passive shielding will indeed further decrease the count rate. This is further supported by the fact that the measured attenuation of e.g. the $^{40}$K and the $^{208}$Tl peaks due to the preliminary shielding leads to very similar effective thicknesses of 13\,cm and 16\,cm, respectively, compared to the actual shielding of 15\,cm. We derive that the final setup with a lead shielding of 20\,cm will lead to background count rates for GePD2 well below 500\,kg$^{-1}$d$^{-1}$, comparable to the world's most sensitive HPGe detectors \cite{Turkat2023}.

While the sensitivity of GePD2 will certainly enable the investigation of multiple transitions during the $^{12}$C+$^{12}$C campaign at the Bellotti IBF, the main focus will however be on the emission of the $E_{\gamma}=440\,$keV and $E_{\gamma}=1634\,$keV photons stemming from the $^{12}$C($^{12}$C,$p$)$^{23}$Na and the $^{12}$C($^{12}$C,$\alpha$)$^{20}$Ne reaction, i.e. from the respective deexcitations of their first excited state to the ground state. For this preliminary setup the background count rates in these ROIs are 3.6(1)\,keV$^{-1}$d$^{-1}$ and 0.44(3)\,keV$^{-1}$d$^{-1}$, and are expected to further improve in the final setup. The expected FWHM of the two ROIs at $E_{\gamma}=440\,$keV and $E_{\gamma}=1634\,$keV are 2.3\,keV and 2.9\,keV, respectively. It is worth mentioning, that these values were determined based on background peaks and do not account for any expected Doppler broadening, which might especially impact the sensitivity for the $^{12}$C($^{12}$C,$\alpha$)$^{20}$Ne reaction.


\subsection{Sensitivity study for $^{12}$C+$^{12}$C}

To perform a sensitivity study, both the reaction count rate and the background count rate (as well as its standard deviation) need to be known. However, due to the wide range of theoretical predictions regarding the $^{12}$C+$^{12}$C reaction cross section towards lower energies, this sensitivity study needs to be treated with caution. While Tumino \textit{et al.} \cite{Tumino2018} report a comparatively large cross section due to several resonances enhancing the reaction rate at the astrophysically relevant energies, Mukhamedzhanov \textit{et al.} \cite{Mukhamedzhanov2019} claim significantly lower cross sections. None of these extrapolations can yet be clearly excluded or favored based on available data sets. The still widely accepted extrapolation based on experimental data sets in literature provided by Caughlan and Fowler (CF88) \cite{Caughlan1988} lies in between the predictions of Tumino and Mukhamedzhanov. It is emphasized, that the actual sensitivity will strongly depend on the real cross section, which is why all three rates mentioned above are taken into account in the following.

To compare the expected amount of signal events per day in the region of interest with the expected background count rate, several assumptions are made. A $^{12}$C$^{+}$ beam with an electrical beam current of 200$\,\upmu$A is assumed and the full energy peak efficiency is estimated based on 
Geant4 simulations (cf. \autoref{sec:GePD2_Characterization} for a description 
of the simulation tuning).
The background count rate is given for two cases: the currently achieved background level (cf. \autoref{fig:GePD2_Spectrum}), as well as the background level that can be expected in case of the final setup \cite{Caciolli2009}.

The sensitivity limit is calculated considering a run of 30\,days of irradiation, which is the maximum time we foresee to acquire data at one energy.
The additional contribution to the background rate due to Compton continua from higher energy \isotope[12]{C}+\isotope[12]{C} reactions is estimated to be $0.4$\% using a Geant4 simulation based on branching ratios listed in \cite{Becker1981}, and was omitted because it is negligible with respect to the environmental background.
Any additional background rate based on in-beam contributions (i.e. due to $^{1}$H($^{12}$C,$\gamma$)$^{13}$N and $^{2}$H($^{12}$C,$p\gamma$)$^{13}$C) is also omitted based on their expected insignificance with respect to the dominant environmental background. This approach is supported based on a dedicated study at the Felsenkeller shallow-underground lab (cf. \cite{Szucs2019,Bemmerer2024}) for the upcoming $^{12}$C+$^{12}$C campaign at Bellotti IBF, as well as investigations during the irradiation of similar targets, where deuterium and hydrogen contamination was reported to be negligible after 20\,minutes of irradiation due to outgasing (cf. \cite{Spillane2007}).

\begin{figure}[tbh!]
\includegraphics[width=\columnwidth]{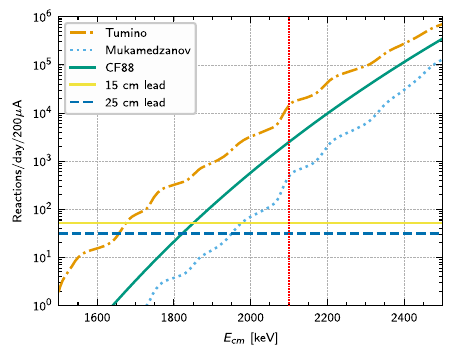}
\caption{\label{fig:GePD2_Sensitivity} Expected number of $^{12}$C($^{12}$C,$p$)$^{23}$Na reactions per day assuming 200$\,\upmu$A and three different reported extrapolations for the cross section mentioned above \cite{Tumino2018, Mukhamedzhanov2019, Caughlan1988}. The detection sensitivity for measuring the signal at 440\,keV with 50\% uncertainty on a $3\sigma$ confidence interval is shown as blue and yellow horizontal lines for the present and final shielding configurations, respectively. The current lower limit for experimental investigations is indicated by a red vertical line.}
\end{figure}

Based on these assumptions, \autoref{fig:GePD2_Sensitivity} shows the resulting reaction rate for the three different reported extrapolations for the cross section (cf. \cite{Tumino2018, Mukhamedzhanov2019, Caughlan1988}) in case of the $^{12}$C($^{12}$C,$p$)$^{23}$Na channel with an emitted $\gamma$-ray at 440\,keV assuming a branching based on \cite{Becker1981}. The resulting detection limits (horizontal lines) are given for an aimed statistical uncertainty of 50\,\%, to be claimed with a $3\sigma$ confidence level. We have determined these limits on the basis of Asimov approximation for discovery significances \cite{Cowan2012, Cowan2011}.

Even the experimentally most unfavorable scenario will enable the investigation of the $^{12}$C($^{12}$C,$p$)$^{23}$Na channel below $E_\textrm{cm}=2$\,MeV, and in the best case scenario the upcoming experimental campaign at Bellotti IBF can even acquire data points below $E_{cm}=1.7$\,MeV.

It is worthwhile mentioning that these estimations are conservatively based on the raw background count rate of the GePD2 detector. For the final configuration of the experimental setup, both the GePD2 detector and the target will be surrounded by the NaI(Tl) array, which will act as an additional Compton suppression shield \cite{Szucs2019}. According to Geant4 simulations, this active veto will further reduce the incoherently scattered events in the background by a factor of 2.6 in the ROI around 440\,keV. Any potential contribution to the count rate of GePD2 based on intrinsic contaminations within the NaI(Tl) detector will be thoroughly discussed in \autoref{sec:NaI}).

\subsection{Intrinsic alpha contamination}
\label{sec:GePD2IntrinsicAlpha}

\begin{figure}[tbh!]
\includegraphics[width=\columnwidth,trim=10mm 15mm 10mm 0mm]{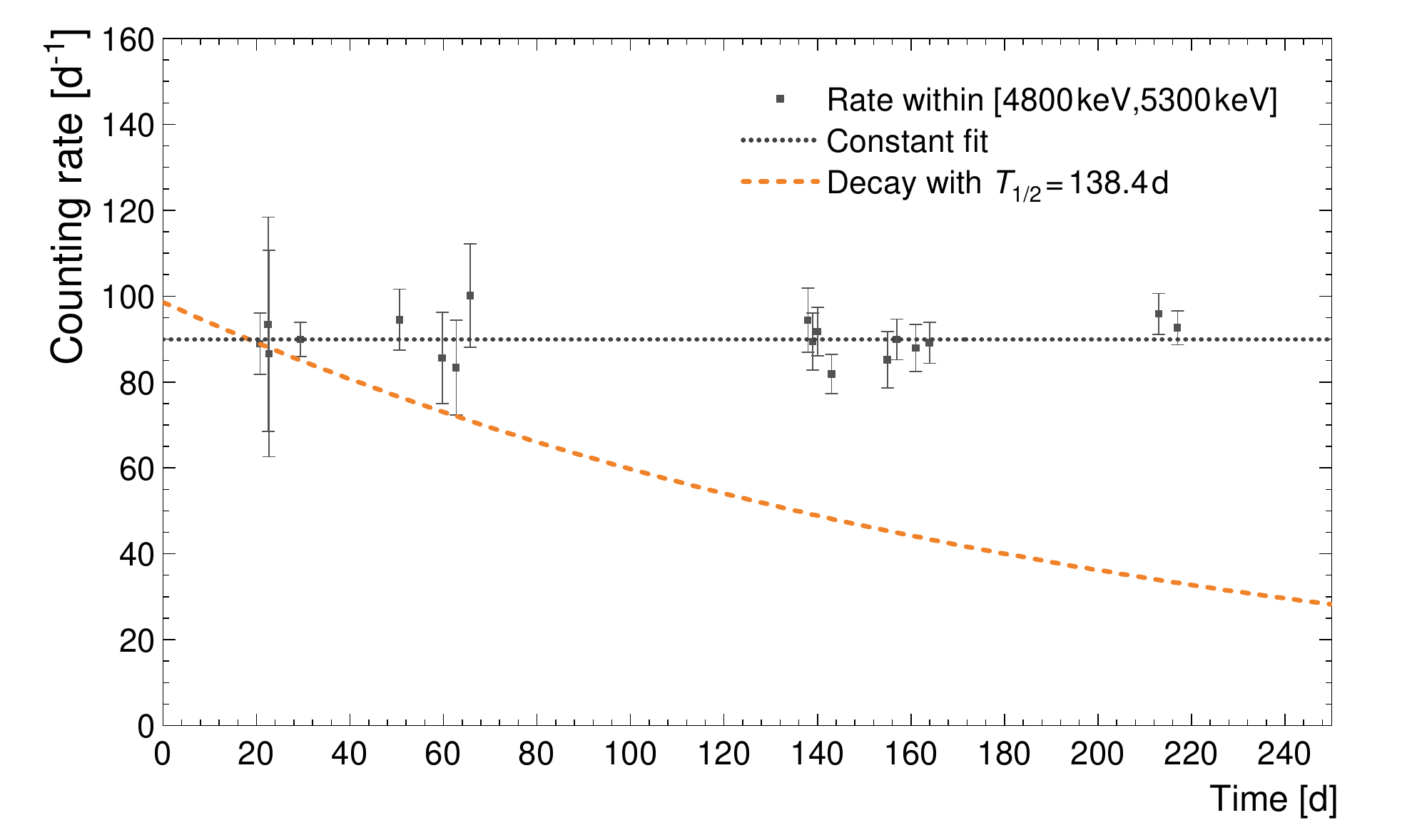}
\caption{\label{fig:GePD2_5MeV_Contamination} Time development of the count rate within an energy interval of [4800\,keV,\,5300\,keV] for GePD2. The black dotted line represents the average of the data set and the orange dashed curve represents the expected trend for a decay which follows the half life of $^{210}$Po, as suggested by \cite{Brodzinski1987}.}
\end{figure}

An apparent particularity of the background spectrum of GePD2 concerns a pronounced and broad peak around $E=5.0-5.2$\,MeV (cf. \autoref{fig:GePD2_Spectrum}). Its count rate is neither influenced by different types of shieldings around the detector, nor did it decrease noticeably within six months of consecutive measurements with respect to its statistical uncertainty (cf. \autoref{fig:GePD2_5MeV_Contamination}). Based on its broad peak shape and left tail its origin is most likely due to a long-lived alpha contaminant located in the vicinity of the crystal.

\begin{figure}[tbh!]
\includegraphics[width=\columnwidth,trim=0mm 0mm 0mm 0mm]{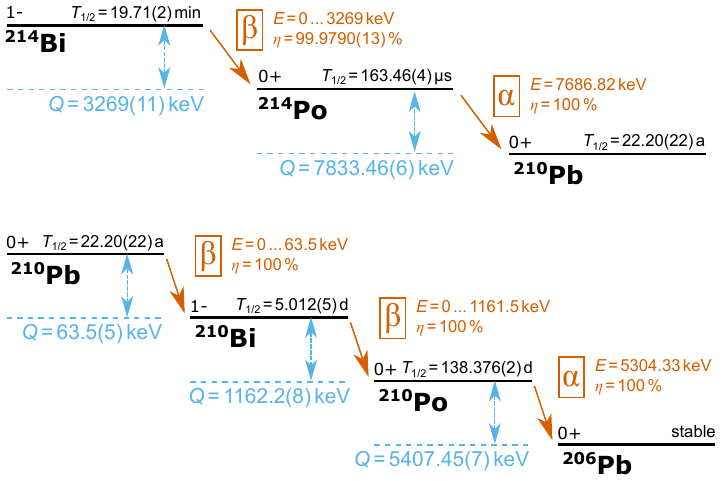}
\caption{\label{fig:DecayScheme}Top: Decay scheme of $^{214}$Bi including the $\beta/\alpha$ coincidence measured within the NaI(Tl) prototype. Bottom: Decay scheme of $^{210}$Pb, which is a long-lived contaminant within GePD2 and feeds the alpha emitter $^{210}$Po. \cite{Kondev2008, Basunia2014, Zhu2021}}
\end{figure}

The most probable origin of these events is a contamination of $^{210}$Pb. It is a beta emitter with a considerably long half life of $T_{1/2}=22.20(22)$\,y (cf. bottom panel in \autoref{fig:DecayScheme}), which decays via another short-living beta transition to its grand daughter $^{210}$Po. While the comparatively low-energy betas cannot be traced directly in the spectrum of GePD2, its daughter $^{210}$Po, however, is an alpha emitter with $E_\alpha = 5304.33(7)$\,keV and $T_{1/2}=138.376(2)$\,d.

Any feeding from parental nuclides more massive than $^{210}$Pb in this decay chain can be excluded due to their short half life. The next long-lived nuclide would be $^{226}$Ra with $T_{1/2}=1600$\,y. However, no indications of its associated alphas can be found in the spectrum.

While this contamination is usually covered by the comparatively intense continuum of muon-induced events in typical laboratories above ground, it is also seen rarely in similar HPGe setups operated within underground laboratories with an attenuated muon flux \cite{Szucs2019,Brodzinski1987,Szucs2015}. 

In Brodzinski \textit{et al.} (cf. \cite{Brodzinski1987}), a broad peak around $E=5.2$\,MeV is associated to a contamination of long-living $^{210}$Pb within the solder close to the active crystal material. They report a decrease of its count rate that seems to be in agreement with the half-life of $^{210}$Po. They argue that $^{210}$Po might have diffused to the surface of the melted soldering material during the manifacturing process and is therefore not in secular equilibrium with $^{210}$Pb anymore.

As shown in \autoref{fig:GePD2_5MeV_Contamination}, the count rate in GePD2 over a period of more than six months is not compatible with the hypothesis of a decaying $^{210}$Po activity, but rather with a constant count rate. Even the high energetic part of the alpha-induced events recorded within more than one month of statistics is only shifted by approximately 88(10)\,keV with respect to the theoretical alpha energy. Assuming an energy loss of 209\,keV/$\upmu$m in pure germanium (amount of doping nuclei is negligible with respect to the germanium), this results in a dead layer thickness of the order of 0.4\,$\upmu$m. It can therefore be concluded that the contamination is neither located on the top nor on the lateral surface of the crystal, where the lithium diffused n+ contact is more than one order of magnitude thicker. 

Due to the fact that the energetic shift matches with the reported thickness of the boron diffused p+ contact of the borehole (0.3\,$\upmu$m according to the data sheet), we can assume that this $^{210}$Pb contamination is located on the surface or close to the p+ contact. Since the count rate is constant, it is not the case for GePD2 that $^{210}$Po decoupled during the soldering process and moved to the surface. It is in secular equilibrium and therefore the spatial distribution of $^{210}$Po and $^{210}$Pb should be identical. Based on the comparatively narrow peak shape it is unlikely that the contamination is due to $^{210}$Pb in any solder close to the crystal. Already a solder joint thickness of $1\,\upmu$m would flatten out the peak significantly to lower energies. 
In fact, it cannot be excluded that the crystal was exposed to $^{222}$Rn during the manufacturing process, which effectively led to a contamination of $^{210}$Pb on its surface.

In conclusion, due to the fact that the mass, the exact location and the distribution of the contamination remains unresolved, we avoid to claim any specific activity. In any case, the count rate of this peak structure within [4800\,keV,5300\,keV] is $R=90(1)$\,d$^{-1}$. 

\subsection{Geometrical crystal characterization}
\label{sec:GePD2_Characterization}
A Geant4 simulation of the HPGe detector is developed based on the data-sheet provided by the manufacturer. To verify and improve the accuracy of this simulation in reproducing the real response of the detector, a scan of the dead and active volumes 
of the HPGe is performed. The results of this measurement are then compared with a simulated scan and used to fine-tune the implemented detector model.

\begin{figure}[tbh!]
\includegraphics[width=\columnwidth,trim=0mm 4mm 0mm 0mm]{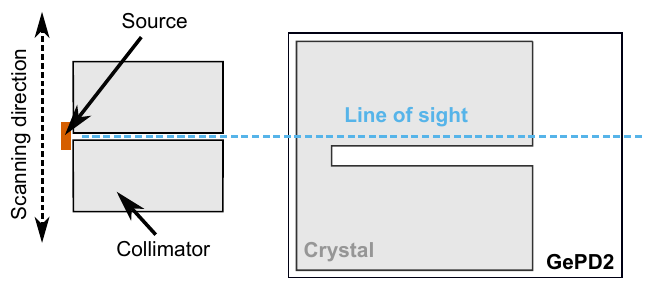}
\caption{Schematic drawing of the setup for the crystal characterization of GePD2. The crystal and the collimator are shown in grey. The source (orange) is placed behind the collimator, which results in a line of sight (blue) for the $\gamma$-rays, respectively. The setup is shown in case of the scan for \autoref{fig:scan_front}.}
\label{fig:SchematicDrawingChraracterizaton}
\end{figure}

To perform the scan, a radioactive source is placed within a movable collimator (cf. \autoref{fig:SchematicDrawingChraracterizaton}).
The collimator is made of a cubic lead brick with a side length of $7.00(5)$\,cm, and a $3.0(2)$\,mm diameter hole
that cuts through its center; the brick 
is mounted in an aluminum frame and moved using a threaded rod and a crank; a ruler is 
fixed on the top side of the frame.
The source is placed behind the $3$\,mm hole, with the detector on the other side of the collimator.
A thin paper sheet is used to electrically insulate the frame from the detector copper housing.
The measurement is conducted using a \isotope[137]{Cs} radioactive source, with an activity of $5.3(1)$\,kBq at the time of measurement.  

The detector is scanned by moving the collimator in fine steps
along its front and side.
Positions and steps are determined based on the HPGe technical drawing to characterize its most prominent features: the dimension of the 
frontal dead layer and the width of the borehole. In particular, the front is scanned 
along the central axis and $3$\,cm above it, 
while the side is scanned along the central axis and vertically 
along the edge of the borehole.

\begin{figure}[tbh!]
\includegraphics[width=\columnwidth,trim=0mm 4mm 0mm 0mm]{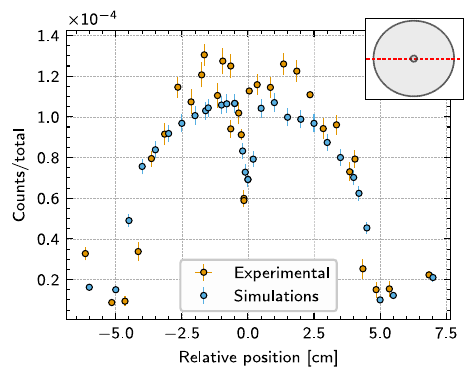}
\caption{Frontal scan of the GePD2 detector along its central axis using a collimated $^{137}$Cs source (see inset).
Experimental data (orange) are compared to results from simulations (blue).
An offset of 0.1 cm was applied to account for misplacement of the collimator.}
\label{fig:scan_front}
\end{figure}

\begin{figure}[tbh!]
\includegraphics[width=\columnwidth,trim=0mm 4mm 0mm 0mm]{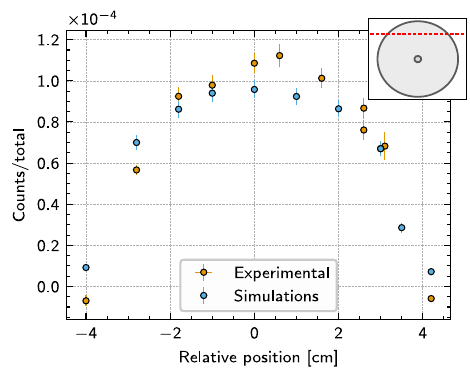}
\caption{Frontal scan of the GePD2 detector along a secant 3\,cm shifted from the central axis using a collimated $^{137}$Cs source (see inset). Experimental data (orange) are compared to results from simulations (blue).
An offset of 0.4 cm was applied to account for misplacement of the collimator.}
\label{fig:scan_frontupper}
\end{figure}

\begin{figure}[tbh!]
\includegraphics[width=\columnwidth,trim=0mm 4mm 0mm 0mm]{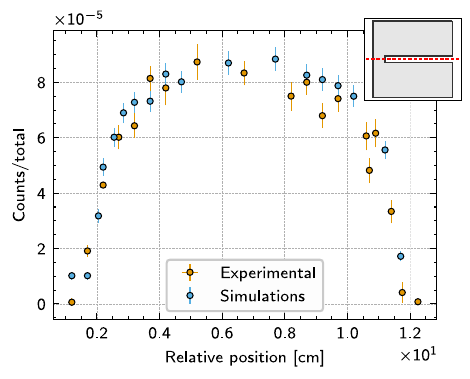}
\caption{Lateral scan of the GePD2 detector along the bore hole axis using a collimated $^{137}$Cs source (see inset). Experimental data (orange) are compared to results from simulations (blue).
An offset of 0.5 cm was applied to account for misplacement of the collimator.}
\label{fig:scan_lateral}
\end{figure}

\begin{figure}[tbh!]
\includegraphics[width=\columnwidth,trim=0mm 4mm 0mm 0mm]{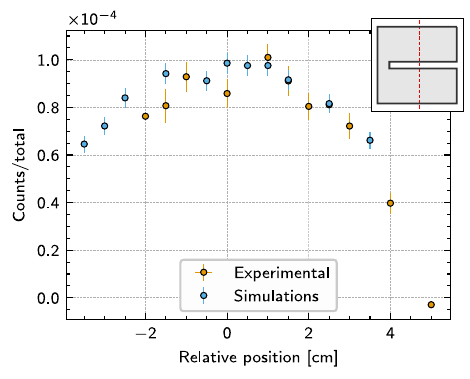}
\caption{Lateral scan of the GePD2 detector perpendicular to the bore hole axis along the center of the crystal using a collimated $^{137}$Cs source (see inset).
Experimental data (orange) are compared to results from simulations (blue).
An offset of 2 cm was applied to account for misplacement of the collimator.}
\label{fig:scan_sidevertical}
\end{figure}

The comparison with the simulation shows that the technical drawing accurately describes the detector. 
Nonetheless some fine-tuning on the dead-layer dimensions and 
size of the borehole is needed:
the length of the borehole is increased by 5\,mm and its diameter 
reduced by 3\,mm, both within the uncertainties provided by the manufacturer. 
As shown in \autoref{fig:scan_front} to \autoref{fig:scan_sidevertical},
after these modifications the simulation reproduces 
the experimental data, with some leftover discrepancies.
These can potentially be associated to a displacement of the position of the 
collimator which happened at mid-measurement.
The setup did not allow for a higher precision investigation.
The results are nonetheless sufficient for a description
of the general shape of the crystal.
Further investigation is needed to address the remaining discrepancies.

\section{The NaI(Tl) detector array}
\label{sec:NaI}

The utilized NaI(Tl) detector from Scionix is a standard low-activity scintillation crystal read out by a SiPM (S14161-6050HS-04 from Hamamatsu) and a built-in preamplifier. The crystal has dimensions of (102\,x\,102\,x\,203)\,mm and is wrapped with a reflector. In direct proximity to the crystal, a semi-flexible optical coupling and a Quartz window are used to guide the light to the SiPM. The surrounding enclosure consists of 1\,mm aluminium and has outer dimensions of (109\,x\,109\,x\,249)\,mm. 

This detector acts as a prototype for the upcoming $^{12}$C+$^{12}$C campaign, where a 16-fold segmented cylindrical NaI(Tl) crystal with the same properties will be utilized as a veto detector surrounding both the target and GePD2. This prototype was therefore tested thoroughly, particularly regarding its intrinsic background contributions since these could potentially influence the sensitivity of GePD2 in later campaigns within the bespoke lead castle. 

\begin{figure}[tbh!]
\includegraphics[width=\columnwidth,trim=15 25 0 0mm]{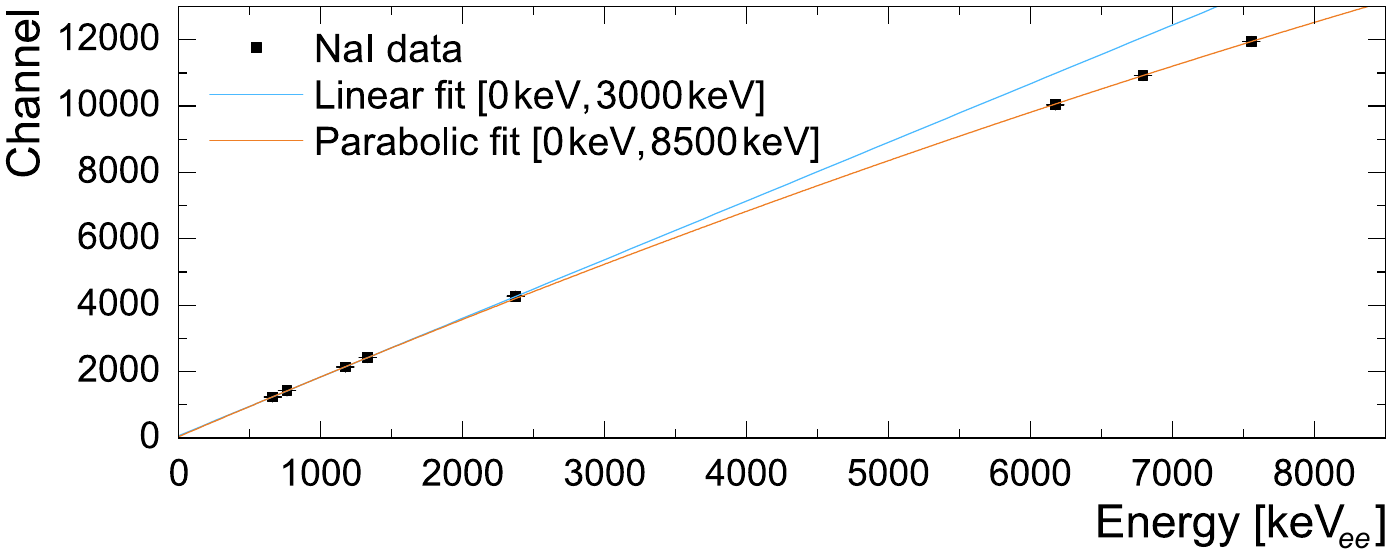}
\caption{\label{fig:NaI_Ecal} Energy calibration of the NaI(Tl) prototype based on $^{60}$Co, $^{137}$Cs and data from the $^{14}$N(p,$\gamma$)$^{15}$O reaction. The entire data set is fitted with a second order polynomial (orange). In addition, a linear fit (blue) is used to solely represent the low-energy data and extrapolated to higher energies for comparability.}
\end{figure}

The light yield in NaI(Tl) detectors is considered to be linear with respect to the deposited energy, as long as same type of particle is considered. Hence, the energy calibration of these detectors is also widely linear, as shown in \autoref{fig:NaI_Ecal} below 3\,MeV$_{ee}$. Based on additional data points using the $^{14}$N(p,$\gamma$)$^{15}$O reaction ($Q=7297$\,keV), it becomes apparent that a second order correction is needed moving towards higher $\gamma$-ray energies. In case of an NaI(Tl) detector equipped with a SiPM, these non-linearities may be mainly associated to energy-dependent light yields, which lead to deviations from the linear behavior towards higher energies, as well as saturation effects in SiPMs, when the number of triggered microcells becomes comparable to the total amount of available cells. We emphasize here that this calibration is only suitable for photon- and electron-induced events and given in terms of electron-equivalent energies (keV$_{ee}$). Any quenching effects will be discussed separately in \autoref{sec:NaI_BG} and \autoref{sec:NaI_Geant4Sim}.

\subsection{Long-term background}
\label{sec:NaI_BG}

\begin{figure}[tbh!]
\includegraphics[width=\columnwidth]{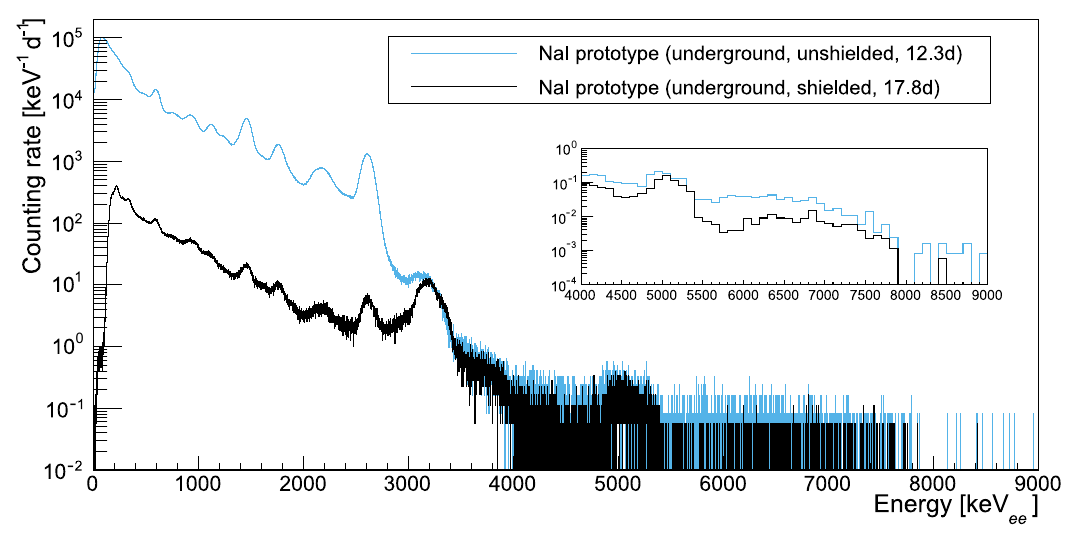}
\caption{\label{fig:NaI_Comparison} Comparison of the count rates for the NaI(Tl) prototype operated underground at LNGS without (blue) and with (black) a surrounding 20\,cm thick lead shield. The high energetic region from 4\,MeV to 9\,MeV is shown as an inset for the same spectra with a binning of 100\,keV.}

\end{figure}

The background spectrum of the unshielded NaI(Tl) prototype (operated underground at LNGS, but outside the Bellotti IBF) shows typical contributions stemming from $^{40}$K and the natural decay chains (cf. blue curve in \autoref{fig:NaI_Comparison}). While this count rate, as expected, significantly drops above 3\,MeV$_{ee}$ due to a lack of contributing $\gamma$-rays, there is an additional contribution around 3\,MeV$_{ee}$, which becomes even more pronounced when an additional 20\,cm lead shield is mounted around the NaI(Tl) prototype (cf. black curve in \autoref{fig:NaI_Comparison}). While the remaining low-energy $\gamma$-ray contributions are strongly attenuated, as expected by this shielding (from 11.8\,x\,10$^{6}$ per day to 8.7\,x\,10$^{4}$ per day in an interval of [300\,keV$_{ee}$,3000\,keV$_{ee}$]), the count rate around 3\,MeV$_{ee}$ remains largely unaffected. In addition, the long-term measurement of the shielded prototype also reveals small higher-energetic features around 4\,MeV$_{ee}$ and 5\,MeV$_{ee}$. These effects due to intrinsic contaminations of alpha-emitters will be discussed, investigated and characterized in \autoref{sec:NaI_PSD} and \autoref{sec:NaI_SelfCoincidence}.

Similar high-energy contributions are also reported in recent publications for comparable detectors operated in underground laboratories, i.e. NaI(Tl) setups \cite{Bernabei2008, Adhikari2024} as well as in other scintillation materials \cite{Chemseddine2024}. Depending on the resulting spectral shape and the available statistics, these contributions are usually associated to alpha contaminations within the crystal, on the surface of the crystal, on the surface of the coating material, or within the coating material. Due to the fact that the light yield differs for alpha particles and (photon-induced) electrons in scintillation detectors (quenching), their energy calibrations, as well as energy dependent resolutions vary drastically, which impedes the investigation and characterization of the alpha contributions. 

As reported in literature (cf. also \autoref{sec:NaIComparisonToOthers}), alpha-induced events in NaI(Tl) detectors are usually due to several types of contamination located at multiple positions within, on, or near the crystal with inhomogeneous distributions. These can typically be fed by the natural decay chains, which might in fact be not in secular equilibrium due to the production mechanism of the surrounding materials and the growing procedure of the crystal, which effectively leads to different specific activities for each part of each chain at each of the possible locations. In case of alpha propagation through insensitive layers, an additional shift and broadening of the distribution of deposited energies is inevitable. To increase the complexity even further, even spatial dependencies of the quenching factor within one detector crystal are reported, which lead to several separate quenching factors, a continuum of quenchings, or a combination of both \cite{Adhikari2024}. The aim of this work to develop a general understanding of the overall contamination, rather than to perfectly reproduce each of the alpha contributions.

\subsection{Pulse Shape Discrimination}
\label{sec:NaI_PSD}

\begin{figure}[tbh!]
\includegraphics[width=\columnwidth,trim=0mm 5mm 0mm 0mm]{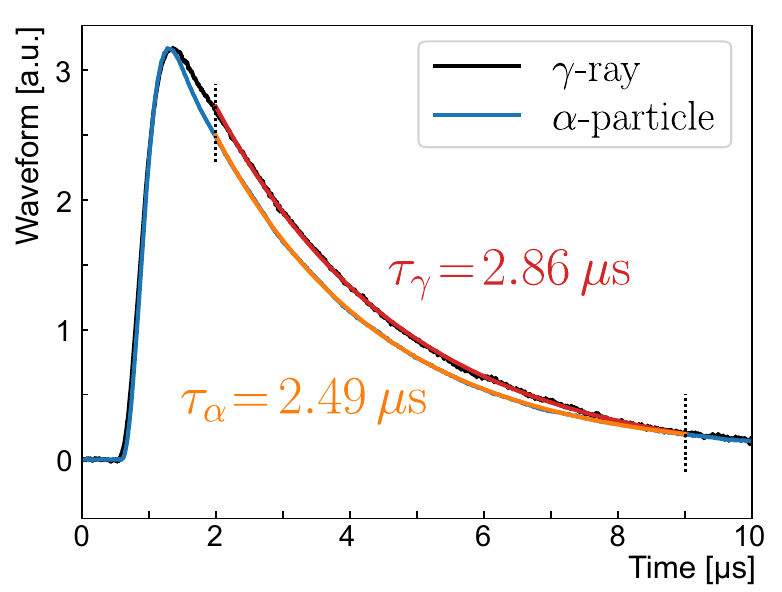}
\caption{\label{fig:NaI_PSD_Fit_Jakub} The observed waveforms for both a $\gamma$-ray (black) and $\alpha$-particle (blue) with the respective fits of the exponential decays in their $[2\,\mu\textup{s},9\,\mu\textup{s}]$ range. The resulting $\tau$s differ by about $0.4\,\mu\textup{s}$, making the decay of the alpha signals faster.}
\end{figure}

\begin{figure}[tbh!]
\includegraphics[width=\columnwidth,trim=0mm 0mm 0mm 0mm]{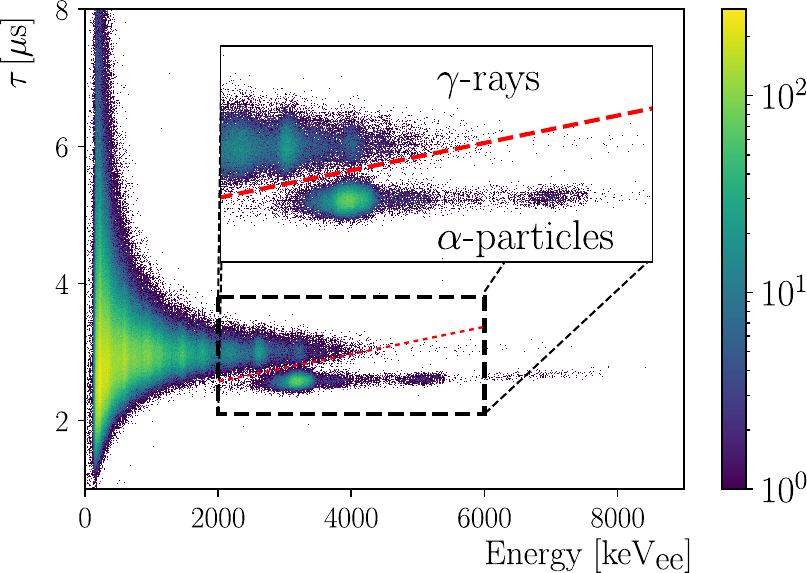}
\includegraphics[width=\columnwidth,trim=5mm 0mm 15mm 0mm]{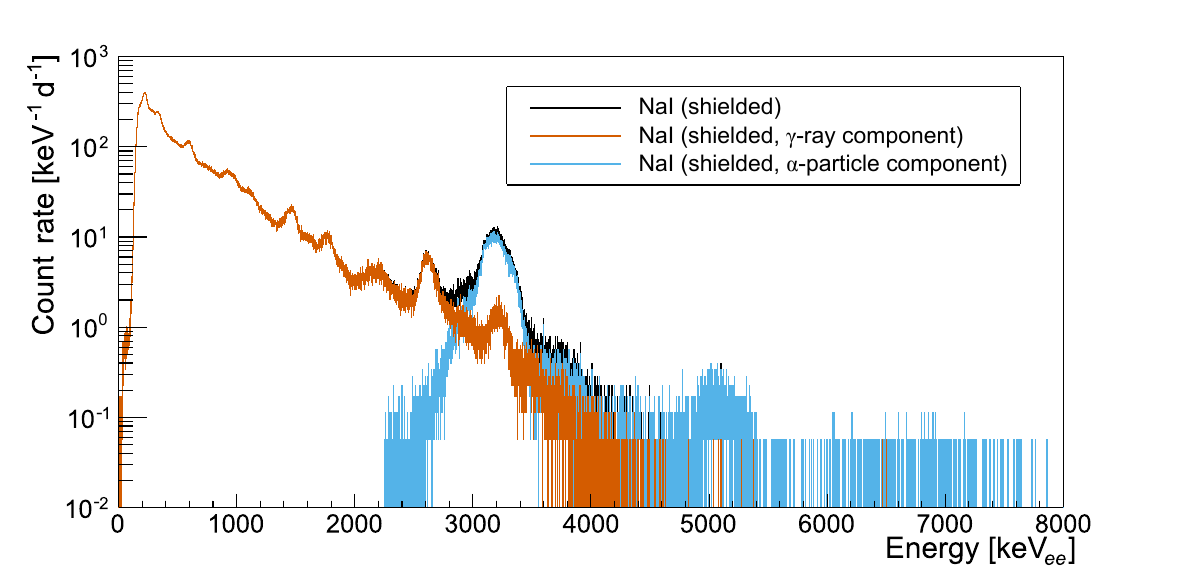}
\caption{\label{fig:NaI_PSD_Jakub}Top: Result of the pulse shape discrimination in case of the shielded NaI(Tl) prototype (cf. \autoref{fig:NaI_Comparison}) showing the decay time of each event with respect to the electron-equivalent energy and the cut that was applied in red dashed line. The amount of events for each bin is color coded as indicated. Bottom: Count rate spectra of the gamma (orange) and alpha contribution (blue) separated according to the analysis shown in the upper panel. The black spectrum represents their sum and is identical to the black spectrum in \autoref{fig:NaI_Comparison}.}
\end{figure}

Due to the presence of both $\gamma$-ray and $\alpha$-particle signals in the energy spectrum, the first goal was to separate the two contributions. The disentanglement would permit us to focus on the latter signal to precisely characterize the internal contamination of the NaI(Tl) crystal, since the $\gamma$-part is largely affected by the environmental radioactivity. Considering that SiPM signals generated by these two contributions differ from each other, an approach similar to \cite{Adhikari2024} was followed. The waveforms of the full signals were acquired, which is different from the HPGe measurements presented above, where a peak height analysis was performed. The sampling rate of the digitizer was 100\,kHz. These waveforms were fitted in the $[2~\mu\textup{s},9~\mu\textup{s}]$ range with an exponential distribution characterized by the decay time, $\tau$. An example of the fit for each of the contributions is shown in \autoref{fig:NaI_PSD_Fit_Jakub}. Once the decay time was obtained for each event, the matrix of these against the associated energies was constructed, which can be seen in \autoref{fig:NaI_PSD_Jakub}. The separation between the $\alpha$-particles and $\gamma$-rays is distinctly visible and permitted us to clearly disentangle the two components based on the faster decay time of the alpha particles.

\subsection{Coincidence analysis}
\label{sec:NaI_SelfCoincidence}

\begin{figure}[tbh!]
\includegraphics[width=\columnwidth,trim=5mm 5mm 5mm 0mm]{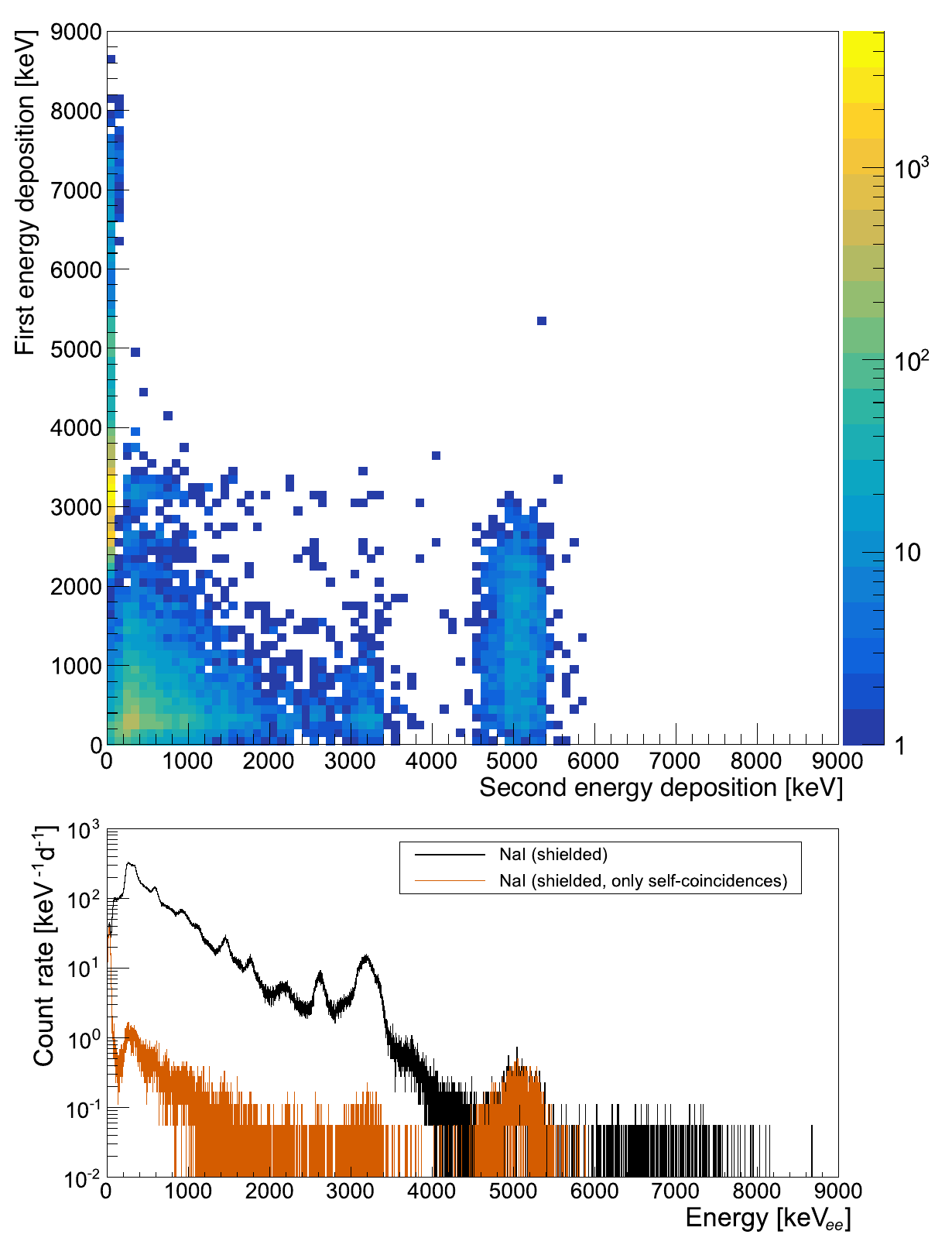}
\caption{\label{fig:NaI_Selfcoincidence}Top: Energy deposition of the first event with respect to the second event in case of two coincident events within the shielded NaI(Tl) prototype. Bottom: Raw count rate spectrum of the NaI(Tl) prototype operated within a lead shielding (black). The latter energy deposition in case of two coincident events within the NaI(Tl) are shown in orange.}

\end{figure}

To further investigate the different contributions within the spectrum, a coincidence analysis was performed searching for subsequently triggered events within a coincidence timing window of $\Delta t_c=2$\,ms of the NaI(Tl) prototype. The result of this analysis is shown in \autoref{fig:NaI_Selfcoincidence} in case of the shielded spectrum (cf. black curve in \autoref{fig:NaI_Comparison}). For all event pairs fulfilling the condition of $\Delta t_c$, the upper panel shows the first deposited energy with respect to the second energy. While the majority of the low-energy events are related to random coincidences due to the comparatively large timing window, there is a clear structure in case of coincident events, where the first events deposit a continuum between $E=0-3\,$MeV$_{ee}$ and the second events subsequently deposit an energy of $E\approx5\,$MeV$_{ee}$.

As shown in \autoref{fig:DecayScheme}, these event pairs can be associated to the beta decay of $^{214}$Bi ($E_{\beta,\textrm{max}}=3.3$\,MeV) with a subsequently occurring alpha decay of $^{214}$Po ($E_\alpha=7.7\,$MeV). Due to the comparatively short half-life of $^{214}$Po ($T_{1/2}\approx163\,\upmu$s), these events can be efficiently triggered as coincidences within the chosen value for $\Delta t_c$. The bottom panel of \autoref{fig:NaI_Selfcoincidence} shows the corresponding histogram of the second triggered events (orange curve) with respect to the raw histogram (black curve), thus providing a distinct value within both the energy calibration, as well as the resolution of the quenched alpha contributions.

\begin{figure}[tbh!]
\includegraphics[width=\columnwidth,trim=10mm 10mm 5mm 0mm]{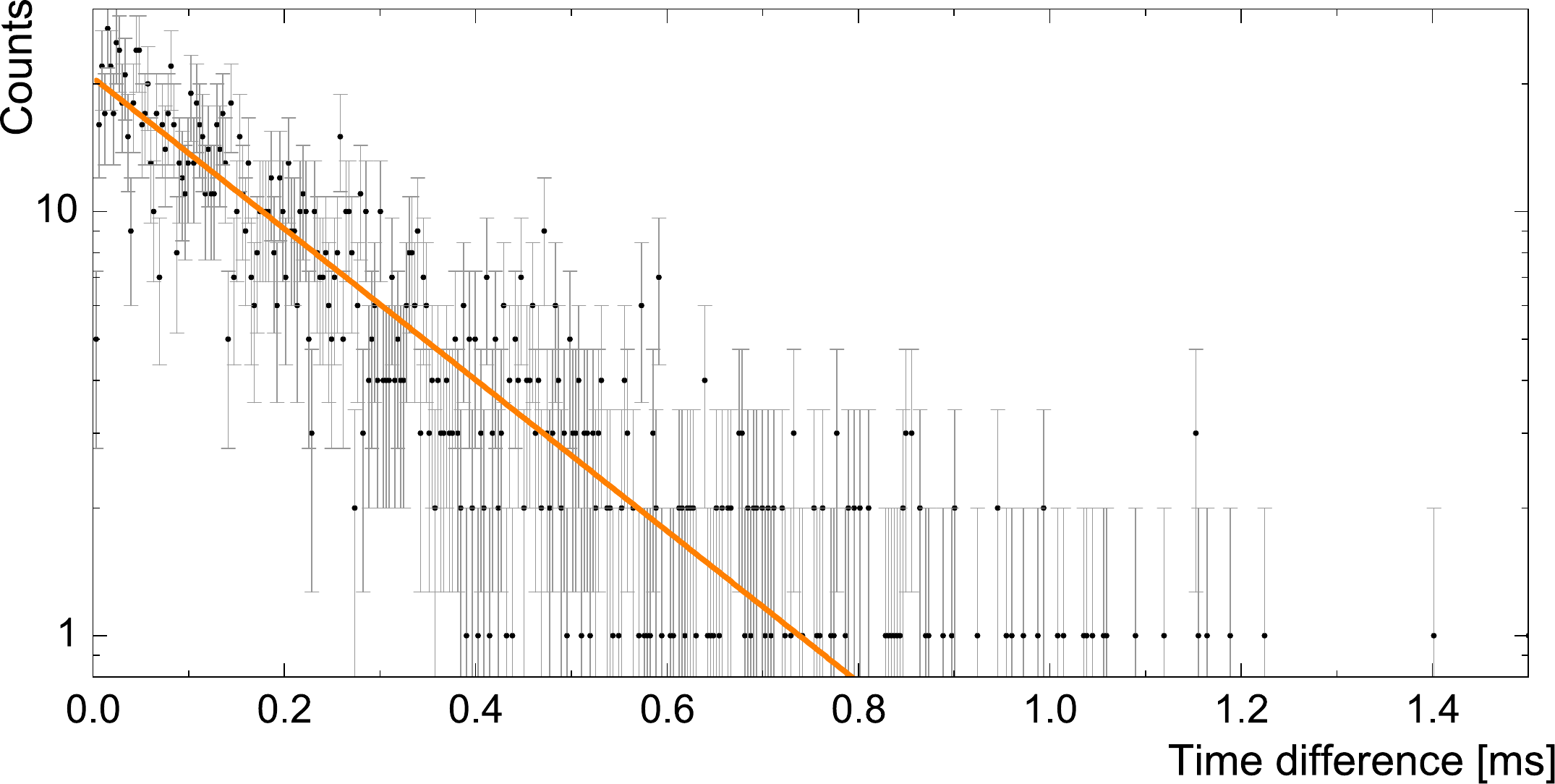}
\caption{\label{fig:NaI_214Po_TimeDiff} Histogram of time differences between coincident events in \autoref{fig:NaI_Selfcoincidence} for events with second energy $>4200$\,keV. The half-life from the fit is $T_{1/2}$=169(5)\,$\upmu$s with the literature value of the $^{214}$Po half-life being 163.46(4)$\upmu$s.
}
\end{figure}

In order to verify the hypothesis of these $\beta/\alpha$ coincidences, an analysis of the time difference is conducted for all event pairs within this distinct feature in \autoref{fig:NaI_Selfcoincidence}. The results are shown in \autoref{fig:NaI_214Po_TimeDiff} and fitted with an exponential function. The corresponding fit value for the half life is $T_{1/2}=169(5)\,\upmu$s, which is in good agreement with the literature half life of $^{214}$Po of $T_{1/2}=163.46(4)\,\upmu$s \cite{Zhu2021}, thus confirming the origin of these coincident events.

\subsection{Geant4 Simulation}
\label{sec:NaI_Geant4Sim}

To characterize the intrinsic radioactivity within the NaI(Tl) prototype, Geant4 simulations were conducted \cite{geant4}. The radionuclides were randomly and homogeneously generated within their respective source medium until stable isotopes were reached. The decay chains were the Geant4 defaults without any alteration. The code can be obtained on GitHub \cite{github}. The pulse shape discrimination (PSD) allows a separate investigation of $\alpha$-particles and $\gamma$-rays, therefore, they are also treated independently within the simulation. To compare the experimental and simulated spectra, the simulations were fitted against the observations with free parameters being the percentages of each component by maximizing the Poissonian log-likelihood with 'scipy' package \cite{scipy}. The results are compared and discussed below.

\subsubsection{Simulation of the alpha contributions\\}
\label{sec:SimAlpha}
As elaborated in \autoref{sec:NaI_BG}, the literature indicates highly complex distributions of alpha contaminations with broken secular equilibria, which are inhomogeneously distributed within or in proximity to the crystal. However, as a first step, a homogeneous distribution of the natural decay chains solely within the crystal was assumed in order to also adjust the energy-dependent quenching function (second order polynomial) and the alpha resolution accordingly.

\begin{figure}[tbh!]
\includegraphics[width=\columnwidth,trim=10mm 15mm 10mm 5mm]{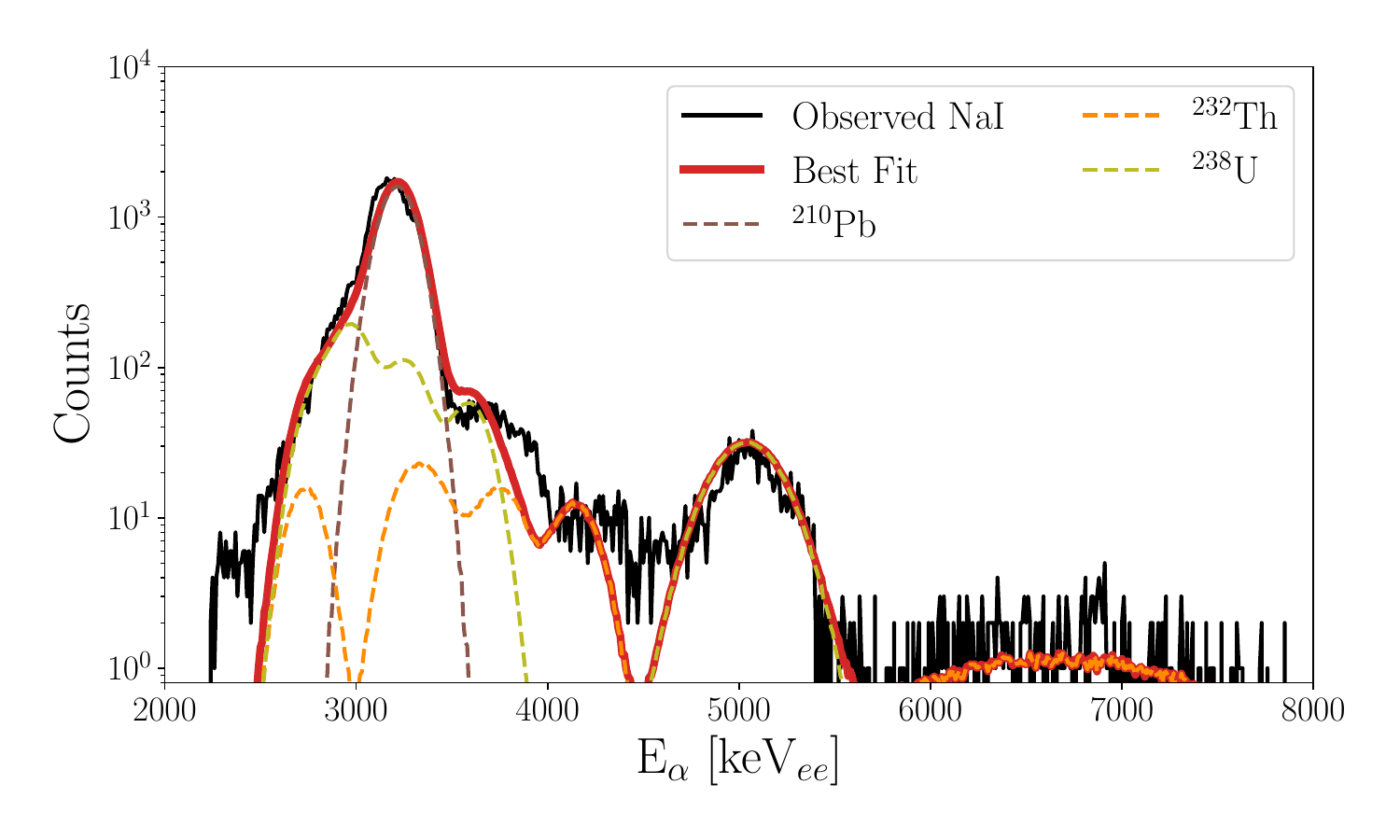}
\caption{\label{fig:NaI_Simulation_Alpha}Comparison between the experimental data of the PSD-separated alpha contributions (black) within the NaI(Tl) prototype and the simulation. The total sum is shown in red, while single contributions are shown in different colors with thinner lines.}
\end{figure}

According to the results of this simulation (cf. \autoref{fig:NaI_Simulation_Alpha}), the primary sources for alpha contributions within the NaI(Tl) prototype are $^{232}$Th and $^{238}$U, as well as their corresponding daughters within their respective decay chains. For this simulation, both chains are assumed to be entirely in secular equilibrium. In addition, a separate contribution from $^{210}$Pb is needed in order to account for the most intense contribution around $E=3200\,$keV$_{ee}$, which is also assumed to be in secular equilibrium with its daughters, i.e. the alpha emitter $^{210}$Po.

The simulation further suggests that the assumption of a homogeneous distribution of these radionuclides within the crystal already provides a sufficient representation of the alpha contributions within the spectrum. In particular, the observed spectral features in the data can be adequately described without accounting for multiple quenching factors, inhomogeneous distributions, or broken decay chains, which are certainly needed for the description of other NaI(Tl) crystals operated in low-background environments. Due to the absence of significant tailing effects in the spectra, significant alpha contributions from the surface of the crystal and from external layers surrounding the crystal can be excluded as well. These results are further discussed in \autoref{sec:NaIComparisonToOthers}.

\subsubsection{Simulation of the gamma contributions\\}
\label{sec:SimGamma}
The main aim for the upcoming section is to phenomenologically understand the remaining gamma contributions within the NaI(Tl) prototype and to investigate their possible origins. Hence, the emphasis is on characterizing the prototype and qualitatively estimating the level of possible major intrinsic contaminations. Therefore, the radionuclides $^{40}$K, $^{210}$Pb, $^{232}$Th and $^{238}$U were homogeneously simulated outside a 15\,cm thick lead castle to compare the resulting spectral shapes, i.e. peak structures and Compton continua to the data.

\begin{figure}[tbh!]
\includegraphics[width=\columnwidth,trim=10mm 15mm 10mm 0mm]{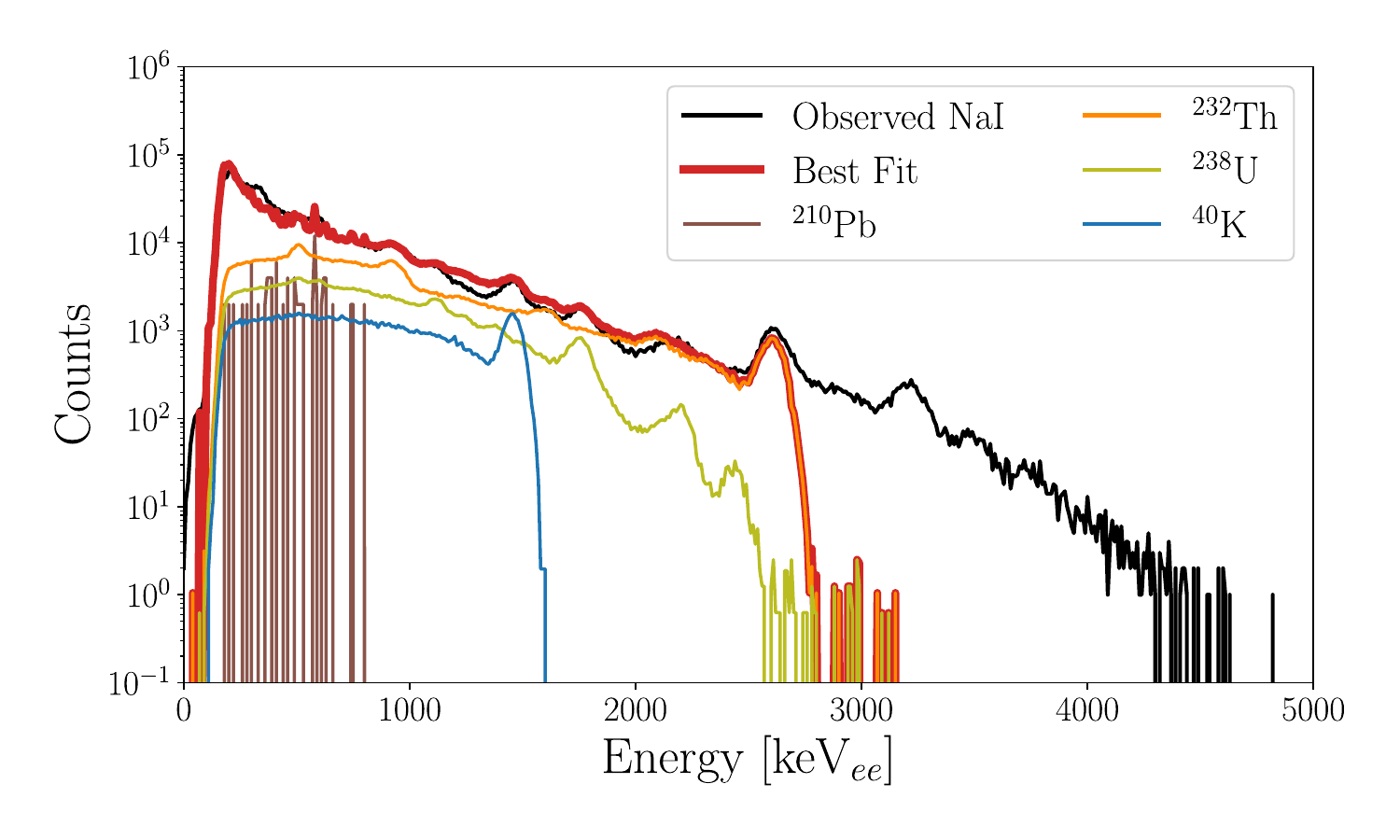}
\caption{\label{fig:NaI_Simulation_Gamma}Comparison between the experimental data of the PSD-separated gamma contributions (black) within the NaI(Tl) prototype and the simulation. The total sum is shown in red and single contributions are shown as well, as indicated in the legend.}
\end{figure}

The resulting fit (cf. \autoref{fig:NaI_Simulation_Gamma}) supports the hypothesis that the majority of gamma-induced events in the NaI(Tl) prototype stem from external contributions outside the lead castle. On the one hand, the thick lead castle impacts the energy-dependent attenuation, which inevitably leads to severe distortions in the ratios between the intensity of low- and high-energy peaks within each decay chain. A dominating contribution stemming from inside the lead castle is therefore not in agreement with the experimental spectral shape. On the other hand, a dedicated simulation was conducted based on the specific activity obtained by the results from the alpha-induced events (cf. \autoref{tab:NaIspecificactivities}). It suggests, that these intrinsic contaminations would only lead to contributions within the gamma-induced part of the NaI(Tl) spectrum, which are more than three orders of magnitude less intense than measured in the experimental data and are therefore considered to be insignificant in the following. Furthermore, their impact on the counting rate of GePD2 and therefore its sensitivity will be further discussed in \autoref{sec:Conclusion}.

As shown in \autoref{fig:NaI_Simulation_Gamma}, the high energy part of the experimental data beyond 2.6\,MeV is not explained by the simulation. This might be due to a convolution of several contributing effects, which are mentioned in the following, but not further investigated due to their insignificance for the upcoming campaign.

These events could be gamma-induced and potentially result from a combination of high-energy, low-intensity gamma-rays above 2.6\,MeV from $^{214}$Bi and $^{208}$Tl. These gammas are not precisely investigated in literature and can therefore lead to discrepancies between the data and the simulation in low-background environments. Another source for contributions from genuine full energy depositions might be due to $^{24}$Na, which can be activated within the crystal, e.g. by $^{23}$Na($n,\gamma$), or other neutron-induced reactions within the crystal material, i.e. $^{127}$I($n,\gamma$)$^{128}$I \cite{Haffke2011}.
Additional contributions might stem from random coincidences, e.g. 1461\,keV + 1461\,keV from $^{40}$K or real coincidences, e.g. 2615\,keV + 583\,keV from $^{208}$Tl.

Furthermore, additional beta contributions from high $Q$-value beta transitions could trigger high-energy events that cannot be disentangled from gamma-induced events in the presented PSD analysis. While some alpha-induced events may potentially also be falsely flagged as gamma-induced events during the PSD (as shown by the slight overlap of both contributions in the upper panel of \autoref{fig:NaI_PSD_Jakub}), this will certainly only account for a negligible fraction. For further insights into high energy contributions in well-shielded NaI(Tl) detectors, see \cite{Ohsumi2002}.

\subsection{Comparison to other NaI(Tl) detectors}
\label{sec:NaIComparisonToOthers}

As emphasized in \autoref{sec:NaI_BG}, the precise investigation of alpha contaminations in scintillation detectors is not trivial due to multiple parameters influencing the overall spectral shape, i.e. location, distribution, non-secular equilibria and (multiple) quenchings. While it is not the main aim of this work, it is however worthwhile comparing the results to available literature, i.e. the identified radionuclides, as well as the resulting hierarchy of specific activities to properly describe their respective NaI(Tl) spectra.

Radiopure NaI(Tl) crystals are typically used in the search of rare-event physics and operated in underground laboratories. Some key experiments operated under these conditions were DAMA/LIBRA, which claimed the observation of an annual modulation due to dark matter \cite{Bernabei2010}, as well as PICO-LON, DM-ICE17 and ANAIS, which were operated mainly to validate this conclusion. These NaI-based experiments are chosen for comparison with our work due to their extensive investigation and detailed reports on specific activities of intrinsic radionuclides. Other large-scale experiments operated under these conditions are KIMS-NaI, COSINE and SABRE. However, they only report upper limits or no activities in case of the considered radionuclides and are therefore not discussed \cite{Adhikari2016, Adhikari2018, DAngelo2016}.

\begin{table}[tbh!]
\setlength{\tabcolsep}{9pt}
\caption{\label{tab:NaIspecificactivities}Intrinsic specific activities of NaI(Tl) crystals used by various collaborations, given in $\upmu$Bq/kg. While this work uses secular equilibria in case of decay chains, some collaborations also report separate values for radionuclides within one chain. Still, only $^{210}$Pb, $^{232}$Th and $^{238}$U are reported here. In case of multiple reported ingots, (b) and (w) indicate the best and worst for those cases where the specific activity of all three radionuclides was determined, respectively.\vspace{3pt}}
\resizebox{\columnwidth}{!}{
\begin{tabular}{l| S[table-format=3.3] | S[table-format=2.3] | S[table-format=2.3] |l}
Experiment     & $^{210}$Pb & $^{232}$Th   & $^{238}$U & Ref. \\
\hline
DAMA/LIBRA & 24.2(16) & 8.5(5)    & 4.4(7) & \cite{Bernabei2008} \\
PICO-LON (b)  & 29(7) & 1.5(19) & \textrm{\textless\,0.5~~~~} & \cite{Fushimi2016} \\
PICO-LON (w)  & 9600(100) & 243(11) & 520(73) & \cite{Fushimi2016} \\
DM-ICE17   & 1500        & 10        & 17 &  \cite{Cherwinka2014} \\
ANAIS (b)   & 700(100)        & 0.7(1)        & 2.7(2) &  \cite{Amare2019} \\
ANAIS (w)   & 3150(100)        & 4(1)        & 10(2) &  \cite{Amare2019} \\
This work  & 3300          & 20           & 110 & 
\end{tabular}
}
\end{table}

As shown in \autoref{tab:NaIspecificactivities}, the overall amount of intrinsic activity varies significantly and certainly depends on the required radiopurity levels for each experiment. However, independent of the overall radiopurity, all reported NaI(Tl) crystals show significantly higher levels of a separate $^{210}$Pb contribution with respect to $^{232}$Th and $^{238}$U. The reason for this separate contribution is most likely an exposure of the powder to $^{222}$Rn prior to crystallization \cite{Adhikari2016}. Furthermore, the reported specific activities of $^{232}$Th and $^{238}$U are usually of the same order of magnitude, with $^{232}$Th typically showing lower specific activities than $^{238}$U. These results are in agreement with ours.

It is worthwhile mentioning that each collaboration treats broken decay chains, multiple quenching factors and necessary places of origin differently for matching their simulations to their experimental result. However, the general trend of the simulations is in sufficient agreement with our results. The remaining deviations between our simulation and our data can most likely be associated to several broken subchains within the $^{238}$U chain, as reported in literature \cite{Bernabei2008, Cherwinka2014}. While higher statistical precision could reveal minor corrections for this model, further refinement is not needed for the upcoming purpose of this detector. 

\section{Conclusion}
\label{sec:Conclusion}

We presented a detailed study of the background characterization for low-background HPGe and NaI(Tl) detectors, which will be utilized for an upcoming experimental campaign to investigate the $^{12}$C+$^{12}$C reaction at the Bellotti IBF. Our results demonstrate the effectiveness of an advanced passive shielding configuration, achieving suppression levels that exceed four orders of magnitude with respect to previous attempts of aboveground experiments in the expected regions of interest. This will be even further improved based on the combination of both detectors and the implementation of the respective active shielding for GePD2 in the final setup configuration. Furthermore, our thorough investigation of intrinsic contaminations within these low-background environments confirmed their required level of radio purity for this experimental study. As supported by simulations, even within this low-background environment, no relevant intrinsic contaminations are expected that could impact the sensitivity. 

While the determined intrinsic activities of natural decay chains within the NaI(Tl) detector might indicate a non-negligible impact on the presented sensitivity study of the contributions to the background counting rate of GePD2 in the final setup, it is worthwhile emphasizing the role of the NaI(Tl) detector as a veto detector within the upcoming campaign. When $\gamma$-rays stemming from these contaminations within the NaI(Tl) detector deposit energy in GePD2, their associated $\alpha$ and $\beta$ particles from the respective decay are expected to also induce a signal in the NaI(Tl) detector and are therefore automatically vetoed out. 

The by far dominating background component will be due to primordial nuclides outside the lead castle. However, this component will be attenuated sufficiently to be well suited for the planned high-precision nuclear astrophysics measurements.

Based on this investigation, a sensitivity analysis was performed supporting the aim to achieve unprecedented experimental insights on the $^{12}$C+$^{12}$C reaction towards astrophysically relevant energies. By achieving such a low-background setup, we anticipate that the upcoming campaign can provide valuable cross section data for the $^{12}$C($^{12}$C,$\alpha$)$^{20}$Ne and $^{12}$C($^{12}$C,$p$)$^{23}$Na channels at energies below $E_{\textrm{cm}}=2$\,MeV even in the most unfavorable scenario. These findings are expected to enhance our understanding of the carbon fusion process, which plays a crucial role in stellar evolution and supernova progenitors.\\

\section{Acknowledgments}
D. Ciccotti, M. Orsini and the technical staff of the LNGS are gratefully acknowledged for their support during setup construction and data taking. Financial support by INFN, the Italian Ministry of Education, University and Research (MIUR) (PRIN2022 CaBS, CUP:E53D230023 and SOCIAL, CUP:I53D23000840006) and through the "Dipartimenti di eccellenza" project "Science of the Universe", the European Union (ERC Consolidator Grant project {\em STARKEY}, no. 615604, ERC-StG SHADES, no. 852016), (ELDAR UKRI ERC StG (EP/X019381/1)) and (ChETEC-INFRA, no. 101008324), Deutsche Forschungsgemeinschaft (DFG, BE 4100-4/1), the Helmholtz Association (ERC-RA-0016), the Hungarian National Research, Development and Innovation Office (NKFIH K134197), the HUN-REN Researcher Mobility Program 2024, the European Collaboration for Science and Technology (COST Action ChETEC, CA16117) and the Hungarian Academy of Sciences via the Lend\"ulet Program LP2023-10. C. G. B., T. C., T. D. and M. A. acknowledge funding by STFC UK (grant no. ST/L005824/1).\\


\bibliographystyle{h-physrev3-etal}
\bibliography{turkat}

\clearpage
\newpage

\end{document}